\documentclass[aps,prl,twocolumn,showpacs,amsmath,amssymb,superscriptaddress]{revtex4}

\usepackage{psfrag}
\usepackage{graphicx}
\usepackage{dcolumn}
\usepackage{bm}
\usepackage[all]{xy}

\begin{document}


\title{Bloch oscillation of a Bose-Einstein condensate in a
  subwavelength optical lattice}

\author{Tobias Salger}
\affiliation{Institut f\"ur Angewandte Physik, Universit\"at Bonn,
  Wegelerstr. 8, D-53115 Bonn}
\author{Gunnar Ritt}
\altaffiliation[Present address: ]{FGAN-FOM, Gutleuthausstra{\ss}e
    1, D-76275 Ettlingen}
\affiliation{Physikalisches Institut, Universit\"at T\"ubingen, Auf
  der Morgenstelle 14, D-72076 T\"ubingen} 
\author{Carsten Geckeler}
\affiliation{Institut f\"ur Angewandte Physik, Universit\"at Bonn, Wegelerstr. 8, D-53115 Bonn}
\affiliation{Physikalisches Institut, Universit\"at T\"ubingen, Auf der Morgenstelle 14, D-72076 T\"ubingen}
\author{Sebastian Kling}
\affiliation{Institut f\"ur Angewandte Physik, Universit\"at Bonn,
  Wegelerstr. 8, D-53115 Bonn}
\author{Martin Weitz}
\affiliation{Institut f\"ur Angewandte Physik, Universit\"at Bonn,
  Wegelerstr. 8, D-53115 Bonn}

\date{\today}
\begin{abstract}
We report on experiments studying transport properties of an atomic
Bose-Einstein condensate in an optical lattice of spatial period
$\lambda/2n$, where $n$ is an integer, realized with the dispersion of
multiphoton Raman transitions. We observe Bloch oscillations, as a
clear effect of quantum transport, in the sub-wavelength scale
periodicity lattice. The unusually large tunneling coupling between
lattice sites is
evident from the measured effective mass. Future prospects of the
novel lattice structures are expected in the search for new quantum
phases in tailored lattice structures up to quantum computing in
optical nanopotentials.
\end{abstract}

\maketitle
Optical lattices have developed into successful model systems for
effects known or predicted in solid state physics \cite{5}. Bloch
oscillations, in which an atom subject to a force performs an
oscillatory rather than a uniformly accelerated motion, are one of the
most striking quantum transport property arising from the periodic
potential \cite{6}. In other works, concepts as number squeezing
\cite{7} or the Mott-insulator transition \cite{8} were
investigated. Further, Landau-Zener transitions have been studied in
optical lattices of variable spatial symmetry \cite{9}. So far, all 
experiments studying quantum transport and exploring the strongly
correlated regime have been carried out in optical lattices with 
$\lambda/2$ or above spatial periodicity, i.e. with a periodicity that
does not beat 
the Rayleigh resolution limit. Conventional optical lattices are formed
by atoms confined in the antinodes of an optical standing wave by
light forces that the $\lambda/2$ spatial periodicity of
the trapping potential is naturally imprinted onto the atomic
wavefunction. 

Motivated by the quest to increase the resolution of
optical microscopy as well as to write smaller lithographic features,
multiphoton and entangled photon techniques have been investigated for
the resolving of subwavelength spatial structures \cite{2,3,4}. In
general, both a n-th order multiphoton process as well as a process
involving n entangled photons can lead to a $n$-fold increase in the
spatial resolution. Other developments yielding an optical resolution
beyond the Rayleigh limit include 4$\pi$- and
STED-microscopy~\cite{23}. Subwavelength periodicity optical
lattices are of interest also in the context of the developping
beamsplitters for atom interferometers with large spatial separation
between the paths~\cite{12,21,22}.

Here we report on the observation of Bloch oscillations of atoms in lattices for
which the lattice periodicity is clearly below the Rayleigh resolution
limit. The small spatial periodicity of the investigated tightly
bound subwavelength lattice leads to an increased Bloch period. Within our
experimental uncertainties, no Bragg diffraction signal is observed
when accelerating atoms in the subwavelength lattice to the first band
edge of a conventional standing wave lattice. The transport signals along with the
determined effective mass in a $\lambda/4$-periodicity subwavelength
lattice are compared to the results obtained with a conventional
lattice of $\lambda/2$ spatial periodicity. Besides a modification in
Bloch period, we also find a striking difference in the effective
atomic masses, which is ascribed to the large tunnel coupling between
sites in the high
spatial periodicity subwavelength lattice.

\begin{figure}
\includegraphics[width=0.95\linewidth]{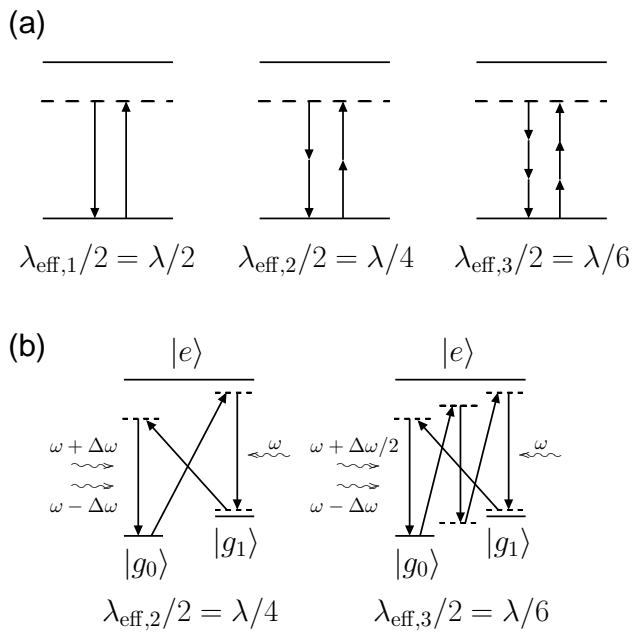}
\caption{\label{FIG.1}Scheme for generation of lattice potentials with
  higher spatial periodicities. (a) Left: Second order processes in a
  usual standing wave lattice, yielding a spatial periodicity of
  $\lambda/2$ of the trapping potential. Middle and right: Four-photon
  (six-photon) processes contributing to a $\lambda/4$ ($\lambda/6$)
  spatial periodicity lattice potential. However, in these simple
  schemes the usual standing wave potential induced by two-photon
  processes dominates. (b) Improved scheme for generation of a
  four-photon (left) and six-photon (right) lattice potential, as used
  in this work. In contrast to the schemes indicated in (a),
  two-photon standing wave processes are here suppressed.}
\end{figure}

Let us begin by describing our scheme to create sub-Rayleigh
resolution optical lattices for cold atoms. The trapping potential of
conventional lattices is determined by the ac-Stark shift in optical
standing waves. In a quantum picture, the absorption of one photon of
a running wave mode followed by the stimulated emission of a photon
into a counterpropagating mode contribute to the trapping potential. A
lattice with spatial periodicity of a fractional harmonic
$\lambda_{{\rm eff},n}/2 = \lambda/2n$ could in principle be achieved
by replacing each of the absorption and emission cycles with a
stimulated multiphoton process induced by $n$ photons, as indicated in
Fig.~\ref{FIG.1}a. Here, $\lambda_{{\rm eff},n}$ denotes the effective
wavelength of a $n$-photon field \cite{10}. However, unwanted standing
wave effects with $\lambda/2$ spatial periodicity also appear in this
process. Fig.~\ref{FIG.1}b shows the used scheme for a four- and
six-photon lattice with potential periodicities of $\lambda/4$ and
$\lambda/6$ correspondingly \cite{9,11,12,13}. Compared to the ladder
scheme, in this improved method, absorption (stimulated emission)
processes have been exchanged by stimulated emission (absorption)
processes of an oppositely directed photon. The high resolution of
Raman spectroscopy between two stable ground states over an excited
state here allows to clearly separate in frequency space the desired
$2n$-order process from lower order contributions.
Our experimental setup has been described previously
\cite{11,14}. Briefly, a rubidium ($^{87}$Rb) Bose-Einstein condensate
is produced all-optically by evaporative cooling in a quasistatic
CO$_2$-laser dipole trap. During the final stages of the evaporation,
a magnetic field gradient is activated, resulting in a spin-polarized
condensate with roughly $10^4$ atoms in the $m_F = -1$ Zeeman
component of the $F = 1$ hyperfine ground state. The lattice beams are
generated by splitting the emitted beam of a tapered diode laser into
two and directing each of the partial beams through an acoustooptic
modulator. The modulators are used for beam switching, and also to
superimpose different optical frequency components onto a single beam
path, as required for generation of the multiphoton potentials with
the schemes shown in Fig.~\ref{FIG.1}b. The beams are send through
optical fibres and focused in a counterpropagating geometry onto the
Bose-Einstein condensate. Respectively to the horizontally oriented
CO$_2$-laser dipole trapping beam, the lattice beams are inclined
under an angle of $41^\circ$ degrees. For the realization of
multiphoton lattices shown in Fig.~\ref{FIG.1}b, we use the $F = 1$
ground state Zeeman components $m_F = -1$ and $0$ as levels
$|g_0\rangle$ and $|g_1\rangle$, and the 5P$_{3/2}$ manifold as the
excited state $|e\rangle$. A magnetic bias field of $1.8$ G removes
the degeneracy of the Zeeman components. For a measurement of Bloch
oscillations, the lattice beams are initially ramped up with a linear
ramp within $20$ $\mu$s to adiabatically load the atoms into the
lowest band of the lattice potential. This procedure was applied for
the usual two-photon as well as for the multiphoton lattice
potentials. One of the lattice beams was subsequently acoustooptically
detuned with a constant chirp rate (for the four-photon lattice scheme
shown on the left hand side of Fig.~\ref{FIG.1}b, the single beam with
frequency $\omega$ was used) to accelerate the lattice respectively to
the atomic rest frame. After a variable acceleration time, the lattice
beams were extinguished and the atomic momentum distribution was
recorded with a time-of-flight absorption imaging technique.
\begin{figure}
\includegraphics[width=0.95\linewidth]{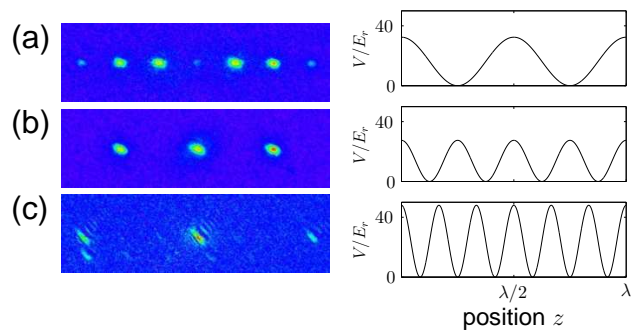}
\caption{\label{FIG.2}(Color online) The figures on the left hand side show 
  far-field diffraction images of a rubidium-Bose-Einstein condensate
  off (a) a usual standing wave lattice with $\lambda/2$ spatial
  periodicity, (b) a four-photon multiphoton lattice with $\lambda/4$
  periodicity and (c) a six-photon lattice with $\lambda/6$
  periodicity. A Stern-Gerlach magnetic field was applied in a angle
  of $45^\circ$ relatively to the horizontal axis, which allows for a
  resolving of the atomic Zeeman structure. The figures on the right
  hand side show the corresponding reconstructed lattice potentials.}
\end{figure}

To verify the effective creation of a subwavelength optical lattice,
we have studied far field diffraction of atoms
off the lattice potentials. The atoms here were exposed to the
periodic potentials imprinted by $6$ $\mu$s long optical pulses of the
lattice beams. Fig.~\ref{FIG.2} shows corresponding atomic
time-of-flight images recorded after a $12$ ms long free expansion
time and reconstructed lattice potentials for (a) a conventional
lattice, (b) a four-photon lattice and (c) a six-photon lattice. For
the higher order multiphoton lattices, the spacing between diffraction
orders is increased due to the smaller spacing between sites of the
corresponding periodic potential. For sufficiently short pulses and
small pulse areas the time-of-flight images are closely connected to
the reciprocal lattice, but also for more general pulses an analysis
of such images allows for a determination of lattice parameters
\cite{10}. The spirit of these preparatory experiments much resembles the atom
diffraction and interferometry work with high momentum transfer of
references~\cite{17,18,19}.
\setlength{\unitlength}{1cm}

Quantum transport of cold atoms in optical lattices much resembles the
behaviour of electrons in crystal lattices \cite{15}. In a
one-dimensional atom potential of the form $V(x) = V_0 \cos^2{(\pi
  x/d)}$, where $d = \lambda/2n$ denotes the lattice periodicity with
n as an integer number and $V_0$ the lattice depth, the energy
spectrum splits up into bands. They can be labelled by the
Eigenenergies $E_j(q)$ of the Eigenstates $|j,q\rangle$, where $j$
denotes the band index and $q$ the atomic quasimomentum, and $E_j(q)$
and $|j,q\rangle$ are periodic functions of the quasimomentum $q$ with
period $2\pi/d = 4\pi n/\lambda$. Conventionally, the
quasimomentum $q$ is restricted to the first Brillouin zone, i. e.:
$|q|\leq\hbar\pi/d = n \hbar k$. Thus, the first Brillouin zone
of a $2n$-th order multiphoton lattice with spatial periodicity
$\lambda/2n$ spans a $n$-fold larger quasimomentum range than a
conventional standing wave lattice of periodicity $\lambda/2$. At the
first band gap, states with quasimomentum $q\in\{-n\hbar k, n\hbar
k\}$ are coupled due to Bragg reflection, which leads to an energy
splitting between the lowest and the first excited band, and similar
couplings also occur between higher bands. When an external force $F$
is applied, the quasimomentum evolves in time and is determined by $q(t) =
q(0) + F t$. At the band gaps we expect, that the wavepackets are
Bragg-reflected, if the force is weak enough, not to cause Landau-Zener
transitions to higher bands, so that the evolution is periodic in time, where $T_B
=n \hbar k/F$ denotes the period, required for the wavepacket to
evolve over the full Brillouin zone. The atomic group velocity
$\langle v\rangle = dE(q(t))/dq$ here oscillates with time.
\begin{figure}
\includegraphics[width=0.95\linewidth]{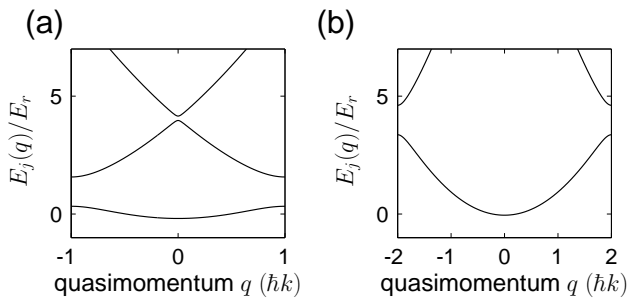}
\caption{\label{FIG.3}Band structure for atoms in a
  periodic lattice potential of (a) $\lambda/2$ and (b) $\lambda/4$
  spatial periodicity respectively. The used lattice depth was
  $2.7\,E_r$ in both cases.}
\end{figure}
The expected band structure for our lattice potentials is shown in
Fig.~\ref{FIG.3} for a two- and four-photon lattice
respectively. 

\begin{figure}
\includegraphics[width=0.95\linewidth]{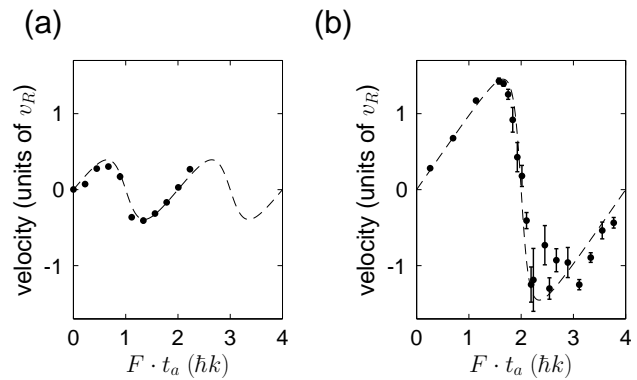}
\caption{\label{FIG.4}Mean atomic velocity as a function of the
  acceleration time $t_a$ for (a) a usual optical lattice and (b) a
  four-photon lattice with $\lambda /4$ spatial periodicity. For the
  data points with no visible error bar the estimated uncertainty
  is below the drawing size of the dots. The dashed line shows a fit
  to the data based on a theoretical model, where the
  Landau-Zener tunneling rate to the next higher Bloch band
  ($\approx$ 10-15\% respectively) was the only free fit parameter.}
\end{figure}

\begin{figure}
\includegraphics[width=0.95\linewidth]{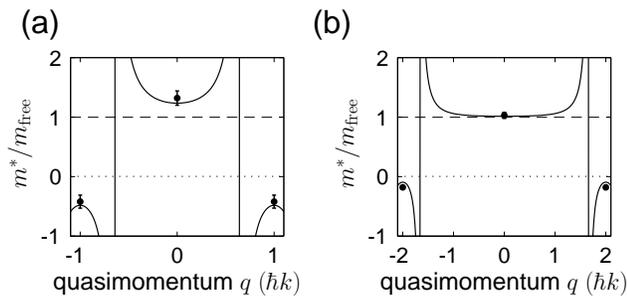}
\caption{\label{FIG.5} Effective mass of atoms in the lowest Bloch
  band as a function of quasimomentum for (a) a usual standing wave
  lattice with $\lambda/2$ spatial periodicity and (b) a four-photon
  lattice with $\lambda/4$ spatial periodicity. The dots with error
  bars are experimental points, derived from the derivative of the
  (interpolated) Bloch oscillation data of Fig.~\ref{FIG.4} for the
  shown specific values of q. The solid lines show the result of a
  theoretical calculation.}
\end{figure}

Experimentally we have adjusted the depth of the lattice
potentials for both lattices to be around $2.7$ $E_r$ (with
$E_r=\hbar^2k^2/2m$ corresponding to the photon recoil energy), as was
monitored by Rabi oscillations \cite{16}. For a measurement of Bloch
oscillations, the atomic Bose-Einstein condensate is adiabatically
loaded into the lowest band of the lattice potentials at zero
quasimomentum ($q = 0$). Subsequently, the lattice potential is
accelerated respectively to the atomic rest frame by applying a linear
variation of one of the Raman beams frequency. In this way, a constant
inertial force $F = -ma$ is exerted onto the atoms, where $m$ is the
atomic mass of the rubidium atoms and $a \approx6.4\mathrm{~m/s^2}$. The
experimental setup is basically the same, as was described in detail
in~\cite{9}. Fig.~\ref{FIG.4}
shows the mean atomic velocity relatively to the lattice as a function
of acceleration time $t_a$ for both two- and four-photon lattice
potentials. For small acceleration times, the atomic velocity
increases linearly with time, as predicted by NewtonÂ´s second law for
a free atom. We expect that an acceleration continues until the edge
of the first Brillouin zone, which is reached at a quasimomentum $q =
\hbar\pi/d$, where $d = \lambda/2n$ denotes the spacing from site to
site. With $n = 1$ and $2$ for two- and four-photon lattices, the band
edge occurs at $q = \hbar k$ and $2\hbar k$ respectively, as was shown in
Fig.~\ref{FIG.3}. For the conventional two-photon lattice, we as in
earlier work \cite{6} observe, that the wavepacket is Bragg reflected
near $F t_a\approx \hbar k$. On the other hand, for the
$\lambda/4$ spatial periodicity four-photon lattice the atoms are
accelerated until $F t_a\approx2\hbar k$ is reached. In both
cases Bragg reflection occurs at the corresponding band gap. The
atomic wavepackets are reflected to the corresponding negative
momentum value and full Bloch oscillations are observed. The
demonstration of this phenomenon for a $\lambda/4$ periodicity
multiphoton lattice directly shows the coherence of atom transport in
such sub-Rayleigh periodicity structures. The Bloch-period $T_B$ in the
smaller periodicity four-photon lattice is a factor two longer than
that of the conventional lattice.
A more detailed comparison of the shapes of the observed oscillations
in Fig.~\ref{FIG.4}, both of which were recorded for comparable
lattice depths, shows, that the slope steepness near the band edge is
larger in the multiphoton lattice than that of the standing wave
potential. Furthermore, the atomic acceleration near zero momentum is
very similar to that of a free atom for the microscopic $\lambda/4$
periodicity lattice, while a somewhat larger difference is observed
for the conventional lattice. These effects can be described in terms
of the effective masses $m^\ast$. The atomic dynamics can be described
using the usual equation of motion $F = m^\ast d\langle v\rangle/dt$
when accounting for an effective mass $m^\ast (q) = 2/(d^2E/dq^2)$,
which in general differs from the mass of a free atom due to the periodic
potential. From our experimental data, we have determined the
effective atomic masses both at $q=0$, and obtained
$m^\ast=(1.29\pm0.12) m_{free}$ and $m^\ast=(1.03\pm0.05) m_{free}$ for
the two- and the four-photon lattice respectively,
and at the band edge at $q=n\hbar k$, which yields
$m^\ast=(-0.42\pm0.11) m_{free}$ and $m^\ast=(-0.18\pm0.02) m_{free}$ for
two- and four-photon lattices respectively. Fig.~\ref{FIG.5} shows
these data points for the effective atomic mass overlayed with the
theoretical prediction (solid lines) as a function of quasimomentum for both
lattices. It turns out, for the smaller periodicity four-photon
lattice, the effective mass near $q = 0$ is
much closer to the real atomic mass than in the usual $\lambda/2$
periodicity lattice, which can be understood in terms of the larger
distance from the band-edge for the multiphoton lattice. If we compare
the tunneling matrix element $J(n)$, defined in the Bose-Hubbard model~\cite{20}
, in the high periodicity lattice with the tunneling rate $J(1)$ in the
case of an optical standing wave of the same potential depth, we
get the following relation: 
\begin{equation}
\frac{J(n)}{J(1)}\propto \sqrt n \exp(-a/n), \nonumber
\end{equation}
where $n$ is an integer and $a$ is a constant number. This formula is
strictly valid in the limit $V_0 \gg n^2 E_r$, where $V_0$ denotes the lattice depth
and $E_r$ the recoil energy, but for the experimentally used
parameters still gives an approximate scaling. One clearly sees the
enhancement of the tunneling matrix element when reducing the distance
$d=\lambda/2n$ between neighbouring lattice sites. A further
interesting issue of our multiphoton lattice is that the ground state
wavefunction size decreases with the small spatial periodicity. We
expect that the effect of interatomic on-site interactions are
enhanced.

To conclude, we have observed Bloch-oscillations of atoms in a novel,
sub-Rayleigh periodicity optical lattice. Evidence for a comparatively
large tunneling coupling between sites in the short periodicity lattice is
obtained from the measured effective atomic mass. We expect that the
observed effects can have applications in the development of nanoscale
quantum computing schemes and the modelling of solid state physics
problems. Note that a reaching of e.g. the Mott-insulator transition
in short-periodicity lattices is favoured by larger tunnelling rates
and stronger interatomic interactions with a decreased spacing from
site to site. An alternative perspective includes the
Fourier-synthesis of arbitrarily shaped lattice structures with
quantum gases realized by superimposing lattices of different spatial
periodicities, which allows for a dynamic tailoring of solid-state
like structures. It would be also of great interest to investigate,
how nonlinear effects, caused by the interaction between atoms in a
Bose-Einstein condensate, are influenced in such a sub-Rayleigh optical
lattice potential. One could for instance, compare the lifetime of
Bloch oscillations in lattices with different spatial periodicities.

\begin{acknowledgments}
We acknowledge financial support from the Deutsche
Forschungsgemeinschaft and the Landesstiftung Baden-W\"urttemberg.
\end{acknowledgments}

\end{document}